\documentclass[conference]{IEEEtran} \IEEEoverridecommandlockouts
\usepackage{cite,amsmath,amssymb,amsfonts,aas_macros}
\usepackage{algorithmic,graphicx,xcolor,textcomp,array,multirow,subcaption}

\definecolor{darkgreen}{RGB}{50,150,0}

\newcommand{\cube}{{\small CUBE}} \newcommand{\piname}{$\pi\,$2.0}
\newcommand{\simname}{{\small Cosmo-$\pi$}}

\def\BibTeX{{\rm B\kern-.05em{\sc i\kern-.025em b}\kern-.08em
    T\kern-.1667em\lower.7ex\hbox{E}\kern-.125emX}}
\begin{document}

\title{ {\tt CUBE} -- Towards an Optimal Scaling of Cosmological
  $N$-body Simulations}


\author{ \IEEEauthorblockN{
    \parbox{\linewidth}{\centering Shenggan
      Cheng\IEEEauthorrefmark{1}, Hao-Ran Yu\IEEEauthorrefmark{2},
      Derek Inman\IEEEauthorrefmark{3}, Qiucheng
      Liao\IEEEauthorrefmark{1}, Qiaoya Wu\IEEEauthorrefmark{2}, James
      Lin\IEEEauthorrefmark{1} }
  }
  \\
  \IEEEauthorblockA{
    \IEEEauthorrefmark{1}Center for High Performance Computing, Shanghai Jiao Tong University, Shanghai, 200240, China \\
    \IEEEauthorrefmark{2}Department of Astronomy, Xiamen University, Xiamen, Fujian 361005, China \\
    \IEEEauthorrefmark{3}Center for Cosmology and Particle Physics, Department of Physics, New York University, New York, 10003, USA \\
    Email: \IEEEauthorrefmark{1}\{chengshenggan, keymorrislane,
    james\}@sjtu.edu.cn,
    \IEEEauthorrefmark{2}haoran@xmu.edu.cn, \\
    \IEEEauthorrefmark{3}derek.inman@nyu.edu,
    \IEEEauthorrefmark{2}wuqiaoya@stu.xmu.edu.cn } }
\maketitle



\begin{abstract}
  $N$-body simulations are essential tools in physical cosmology to
  understand the large-scale structure (LSS) formation of the
  universe. Large-scale simulations with high resolution are
  important for exploring the substructure of universe and for
  determining fundamental physical parameters like neutrino mass.
  However, traditional particle-mesh (PM) based algorithms use considerable amounts of
  memory, which limits the scalability of simulations.
  Therefore, we designed a two-level PM algorithm
  \cube{} towards optimal performance in memory consumption reduction. By using the {\it fixed-point compression} technique, \cube{} reduces the memory
  consumption per $N$-body particle to only 6 bytes, an order of
  magnitude lower than the traditional PM-based algorithms. We scaled \cube{} to 512 nodes (20,480 cores) on an Intel Cascade Lake based supercomputer with
  $\simeq$95\% weak-scaling efficiency. This scaling test was
  performed in \simname{} -- a cosmological LSS simulation using
  $\simeq$4.4 trillion particles, tracing the evolution of the
  universe over $\simeq$13.7 billion years. To our best knowledge, \simname{} is the largest completed cosmological $N$-body simulation. 
  We believe \cube{} has a huge potential to scale on exascale supercomputers for larger simulations. 
\end{abstract}




 \section{Problem Overview}


 In astrophysics, $N$-body simulations are used to study the dynamics
 of globular clusters, galaxy evolutions, galaxy and intergalactic
 interactions and cosmology \cite{1988csup}. Cosmological parameters
 are measured to the percent or even sub-percent levels. This
 precision requires equally accurate numerical modeling of the large
 scale structure (LSS) formation and might lead to cosmological
 $N$-body simulations with a very large $N$.  For example, for studies
 of dark matter and dark energy, we need to resolve the structures and
 assembly histories of faint galaxies in a cosmological volume
 matching modern spectroscopic galaxy surveys. This requires a mass
 resolution of $N$-body particles $M_p\sim 10^8 M_\odot$ and physical
 scale of the simulation $L\sim 1000\,{\rm Mpc}/h$.  Another example
 is the study of neutrino mass using cosmological effects.  To model
 small but nonlinear cosmic neutrino background, we need a large $N$ to
 suppress the Poisson noise (e.g.,\cite{2017NatAs...1E.143Y}).  These
 problems need a dynamical range of 4 to 5 orders of magnitude,
 corresponding to, at least, $10^{12}$ (trillion) particle $N$-body
 simulations. 


 A direct-summation algorithm is unaffordable for a large $N$ because
 the computational complexity of such a pairwise force (PP force)
 calculation is $O(N^2)$. Many algorithms have been designed to reduce
 the complexity to $O(N \log N)$, such as the particle-mesh (PM) or
 tree methods, or even $O(N)$, such as the fast multipole
 method. Fortunately, cosmological $N$-body simulations rarely require
 accurate trajectories of individual particles, but rather just
 correct statistical distributions.  In this scenario, the PM-based algorithms
 usually meet the accuracy requirements and have a potential to
 simulate with a very large $N$.





 However, achieving the largest $N$ with the PM-based algorithms is
 typically bound by memory capacity, instead of computing
 capacity. For example, TianNu \cite{2017NatAs...1E.143Y}, one of the
 world's largest $N$-body simulations, used the P$^3$M algorithm
 \cite{2013MNRAS.436..540H} to complete an $N\simeq 3\times 10^{12}$
 cosmological $N$-body simulation on the Tianhe-2 supercomputer. The
 simulation used all the memory of Tianhe-2, but only 30\% of its
 computing resource\footnote{TianNu used only CPUs in Tianhe-2 which
   contribute to about 30\% of the computing resource.  The rest are
   contributed from the coprocessors (Intel Knights Corner, KNC) which
   associate with small, independent memory, and TianNu did not use.}.
 The challenge to further increase the scale of $N$-body simulations
 is to reduce its memory consumption.



 To tackle this challenge, we designed a two-level PM (PMPM) 
 algorithm \cube{} \cite{2018ApJS..237...24Y} to obtain the lowest memory consumption
 possible. $N$-body simulations typically require
 considerable amounts of memory to store the positions and velocities
 of $N$-body particles. These six numbers occupy 24/48 bytes
 per particle (\texttt{bpp}) if they are stored as single/double precision
 floating numbers. By using an information-optimized algorithm,
 fixed-point format can be used to store the phase-space information
 of particles in memory.  This fixed-point compression can significantly minimize the memory footprint toward only 6 \texttt{bpp}, thus break the memory capacity bound for scaling large $N$-body simulations. 



 In this work, we implemented the \cube{} algorithm in \texttt{C} and optimized the \texttt{C} code for performance and scalability. We then scaled the optimized code on the Intel Cascade Lake based supercomputer \piname{}, to 512 nodes (20,480 cores) with $\simeq$95\% weak-scaling efficiency.  This cosmological simulation, \simname{},
 evolved $\simeq$4.4 trillion cold dark matter (CDM) particles from
 redshift\footnote{The cosmological redshift $z$ is an indicator of
   cosmic time in an expanding universe. The scale factor $a$ of the
   universe satisfies $a\propto 1/(1+z)$.} 99 to 0 in a cubic comoving
 volume $L=3,200\,{\rm Mpc}/h$ per side. To our best knowledge, \simname{} is the largest completed cosmological $N$-body simulation.

\section{Application Scenario}\label{sec.challenge}

High-performance $N$-body simulation codes must address many challenges --
maintaining a low memory footprint given a large $N$,
minimizing the communication across computing nodes, reducing and
accelerating the memory accesses to large arrays, and
efficiently using high-performance libraries to speed up
standard calculations such as FFTs. A series of cosmological
$N$-body codes, including {\small ``Moving-PM''}
\cite{1995ApJS..100..269P}, {\small PMFAST}
\cite{2005NewA...10..393M}, {\small CUBEP$^3$M}
\cite{2013MNRAS.436..540H}, and {\small CUBE}
\cite{2018ApJS..237...24Y}, are designed especially for weak-scaling on supercomputers\footnote{They
introduced, for example, the 2-level particle-mesh (PMPM)
algorithm\cite{2005NewA...10..393M}, optimized on cubic
decompositions \cite{2013MNRAS.436..540H} and fixed-point
information optimization technique \cite{2018ApJS..237...24Y}.}.
Our scaling test is based on our continuous development of
{\small CUBE}, aiming for optimizations on all the above aspects.

\cube{} solves the gravitational force using the PMPM algorithm, 
with optional extended-PP force modules for increased accuracy.  
The traditional PM-based algorithm is suboptimal in parallel computing as it 
requires a full resolution parallel FFT. The PMPM algorithm solves this 
problem by splitting the gravitational force $F_G$ into a short-range force
$F_s$ and a long-range force $F_l$, $F_G=F_s+F_l$. 
The short-range force $F_s$ has a cutoff, i.e., $F_s(r\geq r_{\rm cutoff})=0$, 
and the gravity beyond the cutoff is fully provided by the long-range force, 
$F_l(r\geq r_{\rm cutoff})=F_G(r)$.
When $r<r_{\rm cutoff}$, both force components contribute to $F_G$.
In particular, for smaller $r$, $F_G$ is gradually dominated by $F_s$.
The force-matching optimizations under different criteria are discussed
in Ref.\cite{2005NewA...10..393M,2013MNRAS.436..540H}.

For either $F_s$ or $F_l$, the force calculation is a convolution in real
space, equivalent to multiplying the density field with component-wise 
force kernels in Fourier space, 

\begin{eqnarray}\label{eq.PM}
  \tilde F^i_{s/l}(\boldsymbol{k})=\tilde\rho(\boldsymbol{k})
\tilde K^i_{s/l}(\boldsymbol{k}),
\end{eqnarray}

where $\rho$ is the density field obtained by interpolating particles
onto a mesh (mass assignment), $K$ is the force kernel, a tilde
$\tilde{\ }$ indicates that the variable is in Fourier space, 
and $i=1,2,3$ corresponds to three spatial dimensions. 
The PMPM algorithm essentially computes the short-range ($_s$) and 
long-range ($_l$) components of Eq.(\ref{eq.PM}) separately. The two PM
algorithms are applied on two meshes with different resolutions. 

For the long-range force, because $F_l(r\rightarrow 0)=0$
it requires a lower resolution -- $F_l$ is usually computed on a
4-times-coarser mesh (coarse-mesh). 

In contrast, for $F_s$, the short-range PM is computed on full-resolution fine-meshes.
Because $F_s(r>r_{\rm cutoff})=0$, we can divide the simulation volume $L^3$ 
into many cubic sub-volumes $V_{\rm tile}=(L/N_{\rm tile})^3$, called {\it tiles}, 
and $F_s$ on each $V_{\rm tile}$ is computed by a fine-mesh PM on a slightly 
enlarged volume $V_{\rm PM-fine}=(L/N_{\rm tile}+2r_{\rm cutoff})^3$.
So $F_s$ in $L^3$ is done locally using $N_{\rm tile}^3$ local FFTs.

If the PP force is implemented, the gravity between particles within a preset number
of adjacent neighboring fine-cells are computed explicitly, and the short-range 
force kernel $K^i_s$ is further corrected accordingly.

Thus, only the coarse-mesh FFTs require global MPI communications\footnote{Besides 
computing the gravitational force, MPI communications are also used to send/receive particles 
and to synchronize density/velocity fields.}.
The coarse-mesh PM scales as $O(N\log N)$ but with 1/4 of the full-resolution, so it
typically consumes negligible CPU time. 
This makes the PMPM algorithm excellent for weak-scaling problems
as computing time scales like $O(N)$, because an increase of the problem size leads
to a proportional increase of $N_{\rm tile}^3$ when $V_{\rm tile}$ is fixed.
It also dramatically reduces memory footprint requirements 
as the coarse grid has $4^3=64$ times fewer cells.

Computationally, the global simulation volume is first broken down into small cubical
sub-volumes, each of which is assigned to one MPI process.  A second
level of cubical decomposition occurs inside each process, 
where the node sub-volumes are broken into a
number of local volumes called tiles as mentioned above. Inside an MPI process, the tiles are calculated
in turn, and using OpenMP parallelizes the calculations within
the tile.

\section{Solution}\label{sec.solution}

\subsection{Fixed-point Compression}

The PMPM algorithm is intrinsically memory efficient, and the memory consumption is
thus dominated by the phase-space coordinates of particles.
\cube{} is information-optimized and further reduce this memory footprint
by using fix-point formats instead of floating formats.

This format is based on the structure of the coarse-mesh. 
For the $i$-th dimension, instead of using a 4/8-byte floating number storing 
each particle's global coordinate, $x^i\in[0,L)$, 
we store its relative position w.r.t. its parent-cell in the coarse-mesh.
In particular, if $L$ is divided by $N_c$ coarse-cells per dimension,
we further divide each coarse-cell into 256 bins, and for a particle
in the $n^i_c$-th cell, its coordinate in the $i$-th dimension can
be expressed by

\begin{equation}
  \label{xdecomp}
  x^i=x^i_c+\Delta x^i=(n^i_c-1)L/N_c+(m^i+1/2)L/(256 N_c)
\end{equation}

where $x^i_c$ is the left boundary of the cell, and $\Delta x^i$
is the relative position.
$m^i\in\lbrace 0,1,...,255 \rbrace$ can be stored as an 1-byte integer.
To have the correct $n^i_c$, particles' $m^i$ must be stored in memory in a cell-ordered way. A complementary number count of particle numbers 
in this mesh (density field) will give complete information on 
the particle distribution in the mesh.

The particle velocities in each dimension $v^i$ is decomposed as

\begin{equation}
  \label{vdecomp}
  v^i=v^i_c+\Delta v^i=v^i_c(n^i_c)+f(u^i,z,L/N_c)
\end{equation}

where $v^i_c=\langle v^i\rangle$ 
is the velocity field averaged on the $n_c$-th cell, 
and $\Delta v^i$ is each particle's relative velocity w.r.t. $v^i_c$.
Similarly, $u^i$ can be stored as an 1-byte integer, and maps to 
$\Delta v^i$ by a nonlinear function $f(u^i,z,L/N_c)$.
The nonlinearity is due to the fact that the probability distribution of
$\Delta v^i$ is not uniform but is approximated by a Gaussian distribution with
its variance $\sigma^2_\Delta$ linearly predictable as a function of redshift $z$
and smoothing scale $L/N_c$. In practice, $f$ being the form similar to
the inverse error function (the cumulative distribution function of Gaussian)
minimizes the error in the velocity conversion, because it better solves the dominant
slower moving particles. $\sigma^2_\Delta$ becomes nonlinear at lower 
redshifts, and can be directly measured from simulations.

For each particle, $m^i,u^i\,(i=1,2,3)$ are stored by six 1-byte 
integers. $n^i_c$ and $v^i_c$ information are provided by the density and 
velocity fields on the coarse-mesh. Usually each coarse-cell contains large 
number (64 in \simname{}) particles so the auxiliary density and velocity
fields consumes negligible additional memory \cite{2018ApJS..237...24Y}. 
$m^i$ and $u^i$ can either or both stored as 2-byte integers, corresponding
to a more accurate position/velocity storage. More detailed discussions
and results are in Ref.\cite{2018ApJS..237...24Y}.

\subsection{Optimizations}



\subsubsection{Precompute Expensive Functions}
The fixed-point compression dramatically lowers the required memory footprint of $N$-body simulations; however, the data compression and
decompression introduce extra computing costs. 
The profiling results showed 21\% of total elapsed time is spent on them. For example, calculating Eq.(\ref{vdecomp}) 
is very time-consuming due to the expensive math functions in $f$. 
To implement them efficiently, we precomputed their value and stored them into arrays. 
As the number of compression fixed-point representation is limited, 256 for 1-byte
integer, the time and memory cost of the precomputing are negligible.


\subsubsection{Approximate Expensive Functions}
As the fixed-point compression casts a float (32 bits) to just an 1- or 
2-byte integer, the high accuracy of some expensive math functions is redundant. 
Therefore, we use approximate functions to replace them without 
losing the accuracy of final scientific results. For example, to 
accelerate the expensive $\arctan(x)$ function (which is used in
Ref.\cite{2018ApJS..237...24Y} to approximate velocity compression function ), 
we modified approximate function \cite{rajan2006efficient} and expanded 
its domain (Eq. \ref{arctan}). 
This fast approximate function has a maximum absolute error of 0.0038 radians, 
which has no effects on the accuracy of final results.

\begin{equation}
  \label{arctan}
  \arctan(x) \thickapprox \left\{
      \begin{aligned}
          \frac{\pi}{4} x + 0.273 x (1 - |x|) &, |x| \leqslant 1 \\
          \frac{\pi}{2} - \frac{\pi}{4x} - \frac{0.273}{x} (1 - |\frac{1}{x}|)&, x > 1 \\
          -\frac{\pi}{2} - \frac{\pi}{4x} - \frac{0.273}{x} (1 - |\frac{1}{x}|)&, x < -1 \\
      \end{aligned}
      \right.
\end{equation}

\subsubsection{Compute PP Force Kernel in Mixed-precision}
\begin{figure}[t]
  \centering
  \includegraphics[width=0.48\textwidth]{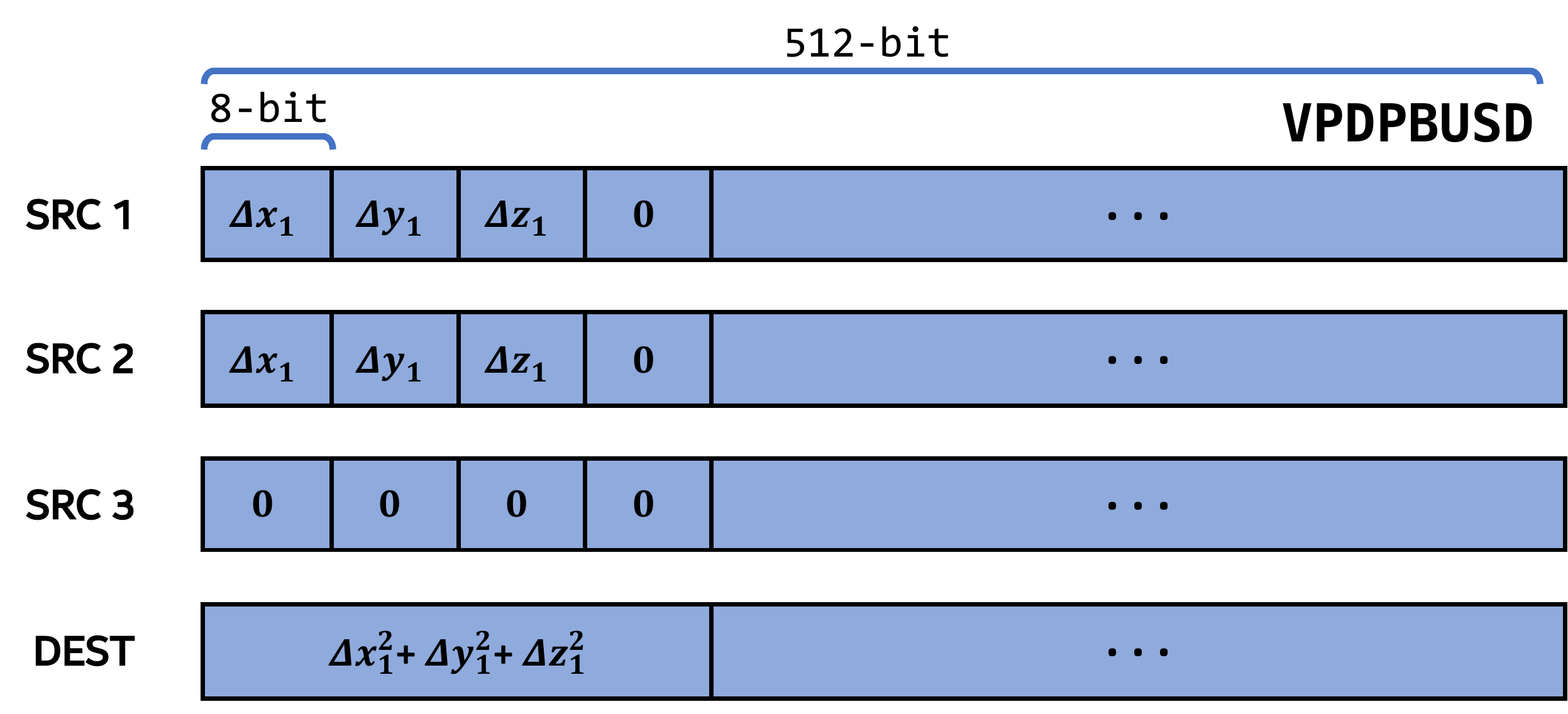}
  \caption{Using Intel AVX512 VNNI instructions to calculate the PP force kernel in mixed-precision. Putting 16 relative vectors of x-y-z axes delta and filling with zero in
    the \texttt{SRC1} and \texttt{SRC2}, then getting the 16 pairs of
    particles with the \texttt{VPDPBUSD} and
    \texttt{VSQRTPS} instructions.}
    
  \label{fig:vnni}
\end{figure}

The PP force kernel requires calculating Euclidean distances between
each pair of particles within a certain range. As the particles positions are compressed in integers, we cannot use traditional AVX512 vectorization instructions. Therefore, we adopt the new AVX512 VNNI (Vector Neural Network Instructions) extension \cite{rod2018vnni} (shipped with Intel Cascade Lake processors and after) to compute the compression position for gravity calculation in mixed-precision. As shown in Fig.~\ref{fig:vnni}, the AVX512 VNNI \texttt{VPDPBUSD} instruction multiplies the
individual bytes (8-bit) of the first source operand by the
corresponding bytes (8-bit) of the second source operand, producing
intermediate word (16-bit) results which are summed and accumulated in
the double word (32-bit) of the destination operand. Theoretically, using the \texttt{VPDPBUSD} instruction can increase 3 times of operations throughput and use 3 times less memory footprint than scalar instructions.






\subsubsection{Distribute MPI Processes}
\begin{figure}[h]
  \centering
  \includegraphics[width=0.48\textwidth]{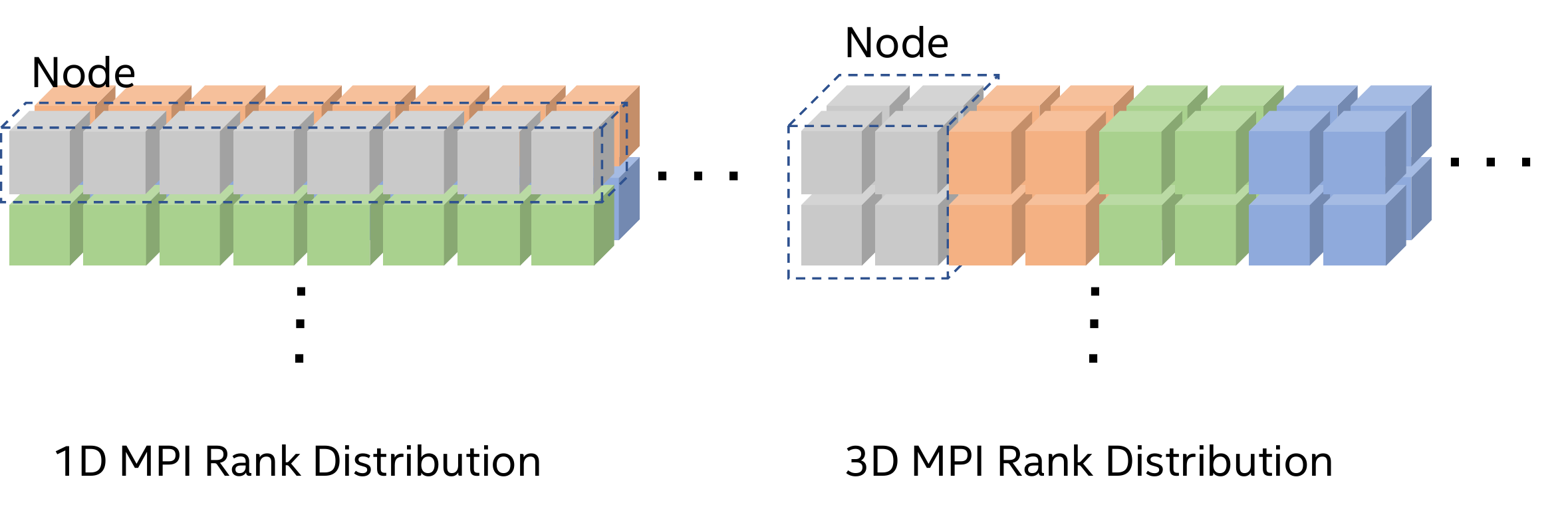}
  \caption{The visual presentation of 1D and 3D MPI Rank Distributions. Each cube represents the physical region for one process. Processes in the same node have
    the same color.}
  \label{fig:mpi}
\end{figure}

The buffer communication requires the positions and velocities of 
particles to be transferred between adjacent grids. To reduce the overhead of buffer communications, as illustrated in Fig.~\ref{fig:mpi}, we use the 3D MPI rank distribution to ensure the processes with adjacent simulation volumes are nearby in the physical domain. 

\section{Performance Metrics and Results}\label{sec.performance}

\subsection{Experimental Setup}

\subsubsection{Hardware \& Software}
We evaluated all scaling tests on supercomputer \piname{}, which has 650 computing nodes with
2PFlops peak performance. Each node has two Intel Cascade Lake 6248 processors (20 cores for each socket) and 192GB DDR4 memory. All nodes are
interconnected with Intel 100Gbps Omni-Path Architecture (OPA).

We compiled CUBE with Intel C/C++ Compiler 18.0.5. The FFT and MPI are supported by Intel MKL 2018 (Update 4) and Intel MPI 2018 (Update 4), respectively. 

\subsubsection{Simulation Parameters}



We use 4,096 MPI processes on 512 nodes ($\sim80\%$ of the full system of \piname{}) to evolve
  $16384^3(\thickapprox 4.39 \times 10^{12})$ cold dark matter
  particles in a $(3.2$ Gpc$/h)^3$ cosmological volume.  We use the
  Zel'dovich Approximation \cite{1970A&A.....5...84Z} to determine the initial
  positions and velocities of particles at redshift $z=99$ and then use \cube{} to evolve
  the particles to $z=0$.
The simulation models a $\Lambda$CDM universe with Hubble
    parameter
    $H_0=100h\ {\rm km}\,{\rm s}^{-1}\,{\rm Mpc}^{-1}=72\ {\rm
      km}\,{\rm s}^{-1}\,{\rm Mpc}^{-1}$,
    CDM density $\Omega_c h^2=0.214 \times 0.72^2= 0.1109$, baryon
    density $\Omega_b h^2=0.044 \times 0.72^2=0.0228$ and initial
    conditions characterized by $\sigma_8=0.80$ and $n_s=$ 0.96. For fixed-point compression, we
  use the 1-byte fixed-point format to store the particle phase
  space.
  
The PM-PM force calculation is computed using a coarse mesh composed of
  $256^3$ cells per process ($4096^3$ in total).  Each process volume is
  decomposed into 8 tiles resolved by $512 + 2 \times 6 = 524$ ($512$ for physical volume and $2 \times 6$ for buffer region)
  fine cells per dimension.  This geometry is equivalent to a
  regular PM calculation with a $16384^3$ fine cell mesh (or $1024^3$ fine cells
  per process), but utilizing 8 times less memory footprint and substantially faster
  due to the decreased size of the global FFT.





\subsection{Performance Metrics}

We choose bytes per particle (\texttt{bpp}) and wall clock time as the two performance metrics for the scaling tests. We use the first metric to measure the memory consumption for each MPI process or each node by using \texttt{/proc/self/statm} and \texttt{MPI\_Allreduce}; we use the second one to measure code speeds by using \texttt{MPI\_Wtime}. 



\subsection{Performance Results}

\subsubsection{Memory Consumption}




Tab.~\ref{tab:memory} lists the memory consumption for one MPI
process of \cube{}, which includes three parts: {\tt Particle}, {\tt
  Coarse mesh}, and {\tt Fine mesh}. {\tt Particle} includes three
components: {\tt Physical region} and {\tt Image buffer} are used
  to store the particle phase space in the physical domain and
buffered region whereas {\tt Tile buffer} is the temporary
buffer for each tile. {\tt Coarse mesh} and {\tt Fine mesh} are the
arrays associated with the coarse mesh and fine mesh, respectively. 
In \cube{}, each MPI process requires 13.75GB memory to simulate $1024^3(=1.074 \times 10^9)$ particles. 
The \texttt{bpp} of \cube{} is thus $13.75 \times 10^9 / (1.074 \times 10^9) =12.8$. 

Due to using fixed-point compression, CUBE has significantly smaller \texttt{bpp} than any other cosmological $N$-body simulation codes. For example, TianNu \cite{2017NatAs...1E.143Y} simulates 2.97 trillion particles on Tianhe-2. Each Tianhe-2 node holds an average of $576^3$ neutrino particles and $288^3$ CDM particles, and uses 40GB memory. Its \texttt{bpp} is thus $(576^3+288^3)/(40 \times 10^9)=186$, which is 14.5 times larger than CUBE's \texttt{bpp}. 



\begin{table}[]
  \centering
  \renewcommand\arraystretch{1.3}
  \caption{Memory consumption for one MPI process to simulate $1024^3$ particles.}
  \label{tab:memory}
  \begin{tabular}{ccrrr}
    \hline
    \multicolumn{2}{c}{\multirow{2}{*}{Parts}}  & \multicolumn{3}{c}{Memory Consumption} \\ \cline{3-5} 
    \multicolumn{2}{c}{}                       & /GB      & /\texttt{bpp}   & Percentage   \\ \hline
    \multirow{3}{*}{Particles}  & Physical region  & 7.3  & 6      & 53\%         \\ 
                                & Image buffer & 1.55     & 2.24   & 11.36\%      \\ 
                                & Tile buffer  & 1.11     & 1.03   & 8.04\%       \\ 
    \multicolumn{2}{l}{Coarse mesh}            & 2.38     & 2.21   & 17.30\%      \\ 
    \multicolumn{2}{l}{Fine mesh}              & 1.32     & 1.22   & 9.56\%       \\ \hline
    \multicolumn{2}{c}{Total}                  & 13.75    & 12.8   & 100.00\%        \\ \hline
  \end{tabular}
\end{table}

\subsubsection{Scalability}

\begin{figure}[b]
  \includegraphics[width=0.5\textwidth]{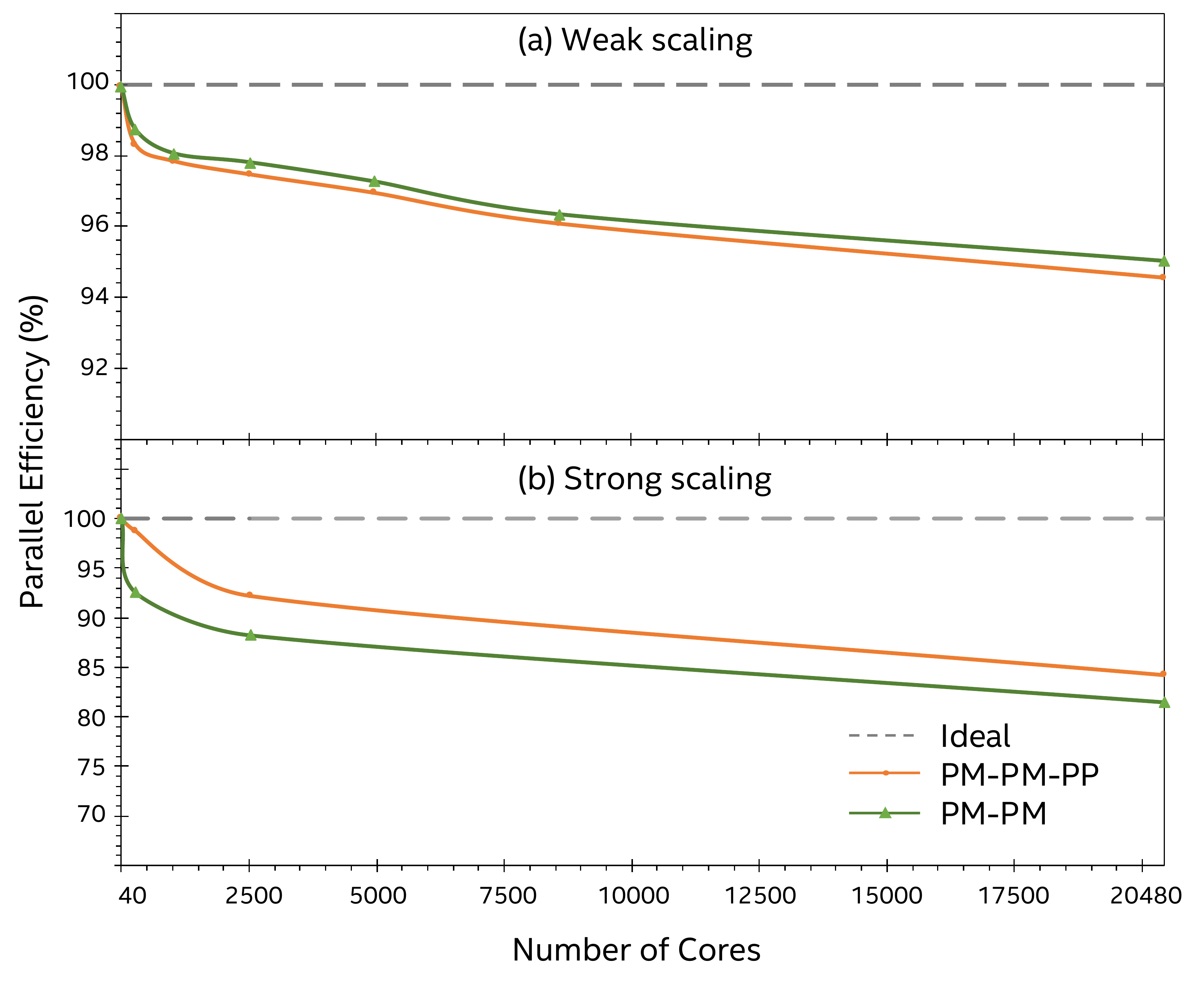}
  \caption{Weak-scaling (a) and strong-scaling (b) from 40 to 20,480
    cores. We show the parallel efficiency
    against core count for PM-PM-PP and PM-PM along with the ideal efficiency.}
  \label{fig:scaling}
\end{figure}

To study the weak-scaling of CUBE, we allow each process to
evolve a $200\ {\rm Mpc}\,h^{-1}$ volume using $1024^3$ fine cells and
gradually scale from 40 cores to 20,480 cores.
Fig.~\ref{fig:scaling}(a) shows CUBE's weak-scaling result both with and
without the PP force (PM-PM-PP and PM-PM in the legend). We see an almost perfect linear speed achieving
95\% parallel efficiency in both cases.  For comparison, the
  TianNu simulation had 72\% weak-scaling efficiency
  \cite{2017RAA....17...85E}; although we note that this scaling test
  was done at redshift $z=5$ where nonlinear structure substantially increases iterations of the PP force kernel.


While often less relevant for cosmological applications, we also
  report CBUE's strong-scaling result. We evolved a fixed global
  simulation volume of $200\ {\rm Mpc}\,h^{-1}$ with $1024^3$ fine cells, but slowly increased the number of cores and distributed corresponding fine cells per process.  Fig.~\ref{fig:scaling}(b) shows the
  strong-scaling where we can see around 81.5\% and 84.2\% parallel
  efficiency at 20,480 cores, respectively.  The degradation in performance with
  increasing cores comes from two aspects: the decrease in computation
  to communication time, and the increase in the ratio of the buffer
  region to physical volume.




\subsection{Validation}
\begin{figure}[h]
  \centering
  \includegraphics[width=0.48\textwidth]{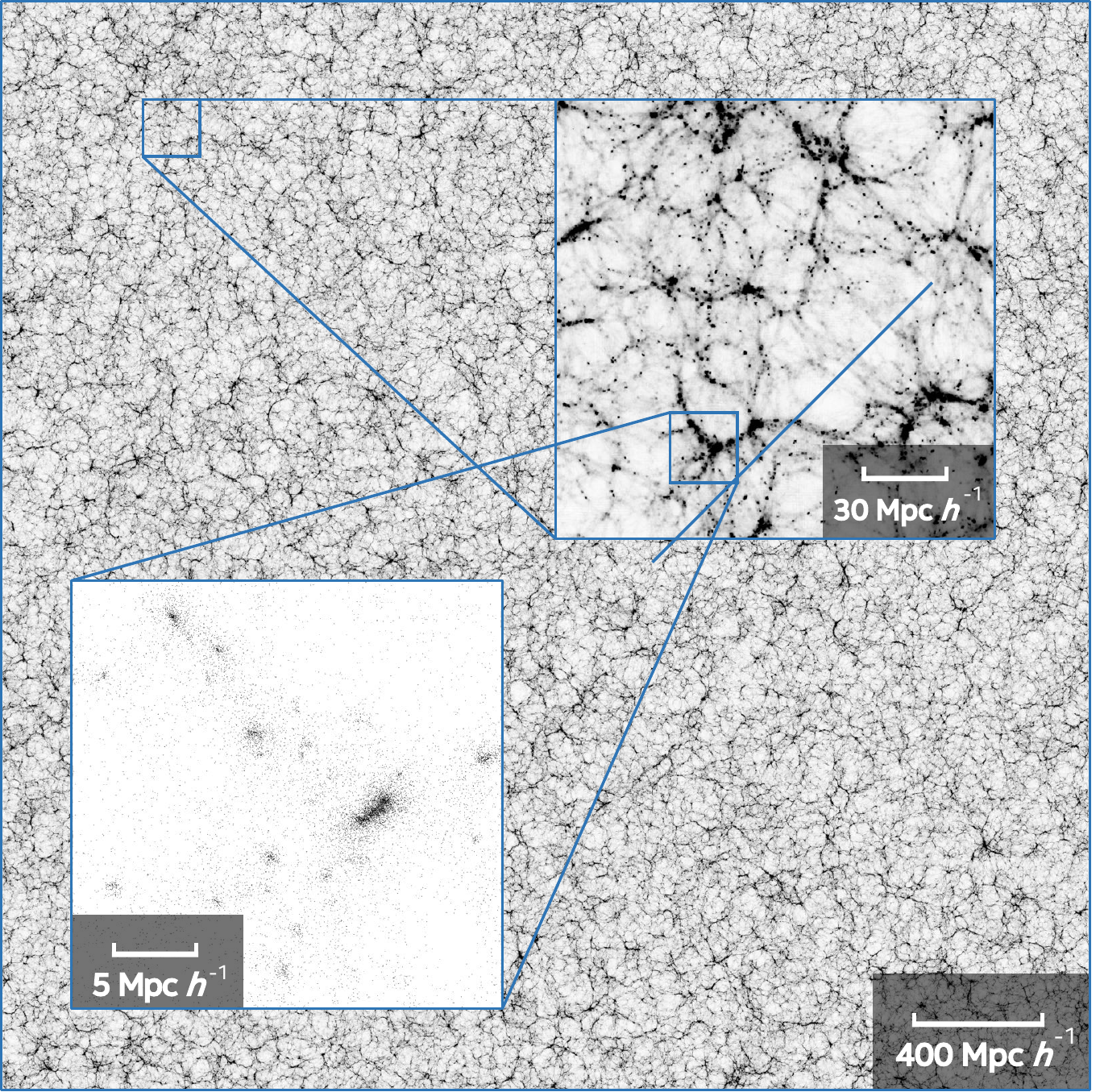}
  \caption{Two-dimensional visualization of the CDM structures in
    \simname{} at redshift $z=0$.  A slice of volume
    $3200\times 3200 \times 20\,({\rm Mpc}\,h^{-1})^3$ is shown,
    while sub-panels show zoomed-in structures.  The high/low column
    densities are rendered by black/white, while the most zoomed-in
    panel shows the direct projection of CDM $N$-body particles.
  }
  \label{fig:cdm}
\end{figure}

We validate the correctness of the \simname{} simulation visually and statistically. 
Fig.~\ref{fig:cdm} illustrates the LSS distribution of CDM in a thin 
slice of the simulation volume. The $N$-body particles in \simname{}
at the final checkpoint $z=0$
are interpolated onto regular grids by the cloud-in-cell (CIC) method.
The resulting 3D density field $\rho$ is then partly projected onto
the plane of Fig.~\ref{fig:cdm} where dark/light colors show high/low
column densities. The size of the figure corresponds to the 
box size $L=3200\,{\rm Mpc}/h$ of the simulation, and the thickness of
the slice is chosen to clearly visualize the structure of the cosmic web --
nodes, filaments, and voids of certain scales.
The hierarchically zoomed-in panels show more detailed CDM structures on 
smaller scales and the most zoomed-in panel shows the projected 
$N$-body particles' distribution directly.

\begin{figure}[h]
  \centering
  \includegraphics[width=0.48\textwidth]{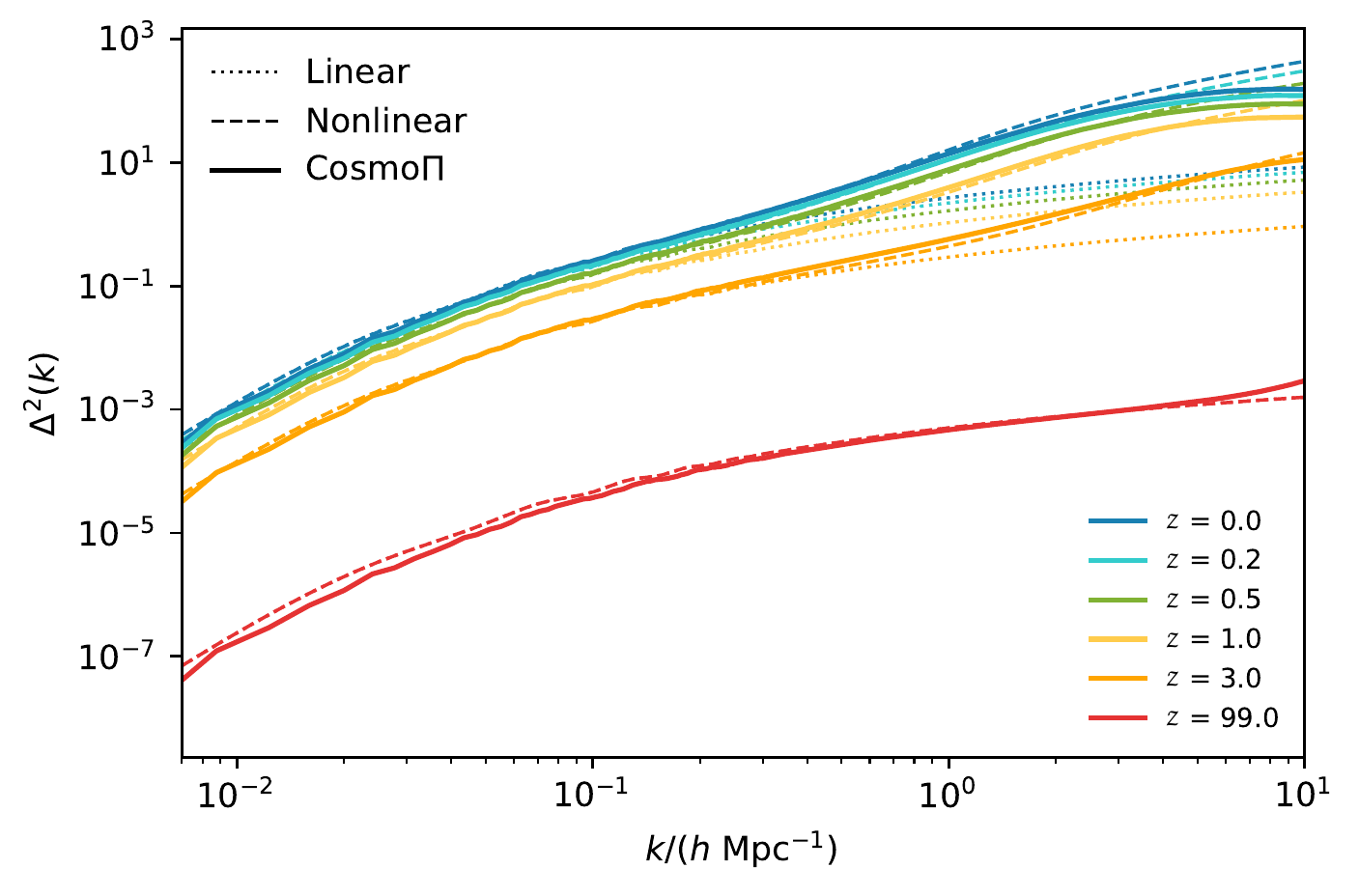}
  \caption{Statistical validation of \simname{}.
    We show the dimensionless power spectra $\Delta^2(k)$ at redshifts
    $z=0,0.2,0.5,1,3,99$ as well as their linear and nonlinear
    predictions. The range of $k$ is chosen to show the transition
    between linear and nonlinear scales.}
  \label{fig:power}
\end{figure}

A useful statistics of LSS is the power spectrum, corresponding to
the Fourier transform of the two-point correlation function.
From the CDM density field $\rho$ we define the dimensionless density 
contrast $\delta\equiv\rho/\bar\rho-1$, from which the
power spectrum $P(k)$ is given by 
$\langle\delta^\dagger({\boldsymbol k})\delta({\boldsymbol k}')\rangle=
(2\pi)^3 P(k){\boldsymbol\delta}_{\rm 3D}({\boldsymbol k}-{\boldsymbol k}')$,
where ${\boldsymbol k}$ is the Fourier wave vector with $k=|{\boldsymbol k}|$ and
${\boldsymbol\delta}_{\rm 3D}$ is the 3D Dirac delta function.
We usually plot the dimensionless power spectrum $\Delta^2(k)$, defined by
$\Delta^2(k)\equiv k^3P(k)/(2\pi^2)$. \simname{} store checkpoints at various 
of redshifts and in Fig.~\ref{fig:power} we plot their dimensionless power 
spectra compared with their linear and nonlinear predictions.
Note that at high redshifts (e.g., the initial condition of \simname{}, $z=99$),
the LSS statistics are well described by linear theory;
while at low redshifts, the linear/nonlinear power spectra\footnote{According to LSS formation
theories, at first order, linear approximations where $|\delta|\ll 1$, 
Fourier modes evolve independently and $P(k)$ is scaled by a growth factor $D(a)$, 
independent of $k$. At lower redshifts, nonlinearities cannot be neglected and we 
use various nonlinear predictions (e.g.,\cite{2011JCAP...07..034B,Smith_2003}) for the power spectrum. 
The exact LSS evolution can only be accurately modeled by $N$-body simulations.}
mismatch and we see $\Delta^2(k)$ follows the nonlinear predictions.
The power spectrum suppression at high wavenumber corresponds to the limited sub-grid
force resolution, which can be improved by using the extended PP force.
The result here is consistent with Ref.\cite{2018ApJS..237...24Y}.

\section{Related Approaches}\label{sec.related}


The gravitational $N$-body simulation is a critical tool for
understanding modern cosmology that relies on supercomputing. The past
ten years has seen the rise of trillion particle simulations starting
with the Gordon-Bell Prize winning simulation \cite{ishiyama20124.45}
and a finalist \cite{habib2012the}. Tab.~\ref{tab:nbody} compares the
large-scale cosmological $N$-body simulations in recent years. There
are now several different codes that have been used to exceed the
trillion particle mark, which differ primarily in their implementation
of the gravitational force computation.  Our work evolved 4.39 trillion particles and thus, to our best knowledge, is the largest cosmological $N$-body simulation. 

\begin{table*}[t]
\centering
\renewcommand\arraystretch{1.3}
\caption{Comparison of the large-scale cosmological $N$-body
  simulations on supercomputers.  For each simulation, we summarize the computational
  scale in CPU cores (c) or GPU cards (g), the problem scale in the particles number ($N_p$), redshift range ($z_i$,$z_f$), box size ($L$), method used to
  compute the force, and force resolution ($\epsilon$). To the best of
  our knowledge, this work is the largest cosmological $N$-body simulation (4.39 trillion particles). 
}
\label{tab:nbody}
\begin{tabular}{c | c | c | c | c | c | c | c | c | c}
  \hline
  Simulations & Years & Codes &Supercomputers & Scale & $N_p (\times 10^{12})$ &
                                                                      $z_i$,
                                                                      $z_f$
  & $L$ (Gpc/$h$) & Force  & $\epsilon$ (kpc/$h$) \\
  \hline\hline
  Dark Sky \cite{skillman2014} & 2014 & 2HOT & Titan & 12,288g \footnotemark[1] & 1.074 & $93,0$ & 8 & Tree & 36.8  \\
  \hline
  $\nu^2$GC \cite{ishiyama2015} & 2015 & GreeM$^3$ & K Computer & 131,072c & 0.55 & $127,0$
  & 1.12 & TreePM & 4.27 \\   
  \hline
  Q Continuum \cite{heitmann2015the} & 2015 & HACC & Titan & 16,384g \footnotemark[1] & 0.55 & $200,0$
  & 0.923  & P$^3$M & 2 \\ 
  \hline
  TianNu \cite{2017NatAs...1E.143Y}  & 2017 & CUBEP$^3$M & Tianhe-2 & 331,776c & $2.97$ & $5,0$ & 1.2 & PM-PM-PP
  & 13 \\
 \hline 
 Euclid Flagship\cite{potter2017} & 2017 & PKDGRAV3 & Piz Daint &
  \textgreater 4,000g \footnotemark[1] & 2.0 & $49,0$& 3 & FMM & 4.8\\
\hline
Outer Rim \cite{heitmann2019} & 2019 & HACC & MIRA & 524,288c & 1.074 & $200,0$ & 3 & TreePM & 2.84\\
\hline
\simname\ (\textbf{This work}) & 2019 & CUBE & \piname{} & 20,480c & \textbf{4.39} & $99,0$ & 3.2 & PM-PM & 195 \\
\hline
\end{tabular} \\
\begin{flushleft}
  \footnotetext[1] 01. These three simulations were carried out using NVIDIA Tesla K20X.
\end{flushleft}
\end{table*}



\section{Impact of the Solution}\label{sec.discussion}


We summarize the impact of the \simname{} simulation in three aspects. First, \simname{} is, to the best of our knowledge, the largest completed
  cosmological $N$-body simulation, evolving 4.39 trillion particles from
  redshift 99 to 0. Simulations such as \simname{}, as well as higher
  resolution ones using the PP force, will allow for improved LSS
  statistics and better understanding of halo assembly and
  substructure.

Second, we believe \cube{} has a huge potential for large-scale cosmological simulations. \simname{} was able to evolve 4.39 trillion particles using just 20,480 cores, a substantial improvement in $N$ enabled by \cube{}'s memory consumption optimization. In the next few years, exascale supercomputers will be available which will allow for simulations using more than ten million cores, increasing the problem scale by at least three orders of magnitude. This will have a profound impact on studies of LSS and other astronomical simulations.

Finally, we show fixed-point compression and mixed-precision calculation can be extremely valuable tools for scientific applications. We expect traditional HPC applications will benefit more from emerging AI computing technologies.

\section*{Acknowledgments}
We thank HPC Center of Shanghai Jiao Tong University for providing 
computing resource and excellent technical support. Hao-Ran Yu acknowledges National Science Foundation of China No.11903021. Shenggan Cheng and Hao-Ran Yu contributed equally to this paper. James Lin is the corresponding author. 

\bibliographystyle{./IEEEtran}
\bibliography{./IEEEabrv,scale_refs}

\begin{thebibliography}{10}
\providecommand{\url}[1]{#1}
\csname url@samestyle\endcsname
\providecommand{\newblock}{\relax}
\providecommand{\bibinfo}[2]{#2}
\providecommand{\BIBentrySTDinterwordspacing}{\spaceskip=0pt\relax}
\providecommand{\BIBentryALTinterwordstretchfactor}{4}
\providecommand{\BIBentryALTinterwordspacing}{\spaceskip=\fontdimen2\font plus
\BIBentryALTinterwordstretchfactor\fontdimen3\font minus
  \fontdimen4\font\relax}
\providecommand{\BIBforeignlanguage}[2]{{%
\expandafter\ifx\csname l@#1\endcsname\relax
\typeout{** WARNING: IEEEtran.bst: No hyphenation pattern has been}%
\typeout{** loaded for the language `#1'. Using the pattern for}%
\typeout{** the default language instead.}%
\else
\language=\csname l@#1\endcsname
\fi
#2}}
\providecommand{\BIBdecl}{\relax}
\BIBdecl

\bibitem{1988csup}
R.~W. {Hockney} and J.~W. {Eastwood}, \emph{{Computer simulation using
  particles}}, 1988.

\bibitem{2017NatAs...1E.143Y}
H.-R. {Yu}, J.~D. {Emberson}, D.~{Inman}, T.-J. {Zhang}, U.-L. {Pen},
  J.~{Harnois-D{\'e}raps}, S.~{Yuan}, H.-Y. {Teng}, H.-M. {Zhu}, X.~{Chen},
  Z.-Z. {Xing}, Y.~{Du}, L.~{Zhang}, Y.~{Lu}, and X.~{Liao}, ``{Differential
  neutrino condensation onto cosmic structure},'' \emph{Nature Astronomy},
  vol.~1, p. 0143, Jul. 2017.

\bibitem{2013MNRAS.436..540H}
J.~{Harnois-D{\'e}raps}, U.-L. {Pen}, I.~T. {Iliev}, H.~{Merz}, J.~D.
  {Emberson}, and V.~{Desjacques}, ``{High-performance P$^{3}$M N-body code:
  CUBEP$^{3}$M},'' \emph{\mnras}, vol. 436, pp. 540--559, Nov. 2013.

\bibitem{2018ApJS..237...24Y}
H.-R. {Yu}, U.-L. {Pen}, and X.~{Wang}, ``{CUBE: An Information-optimized
  Parallel Cosmological N-body Algorithm},'' \emph{The Astrophysical Journal
  Supplement Series}, vol. 237, p.~24, Aug. 2018.

\bibitem{1995ApJS..100..269P}
U.-L. {Pen}, ``{A Linear Moving Adaptive Particle-Mesh N-Body Algorithm},''
  \emph{\apjs}, vol. 100, p. 269, Sep. 1995.

\bibitem{2005NewA...10..393M}
H.~{Merz}, U.-L. {Pen}, and H.~{Trac}, ``{Towards optimal parallel PM N-body
  codes: PMFAST},'' \emph{\na}, vol.~10, pp. 393--407, Apr. 2005.

\bibitem{rajan2006efficient}
S.~Rajan, S.~Wang, R.~Inkol, and A.~Joyal, ``Efficient approximations for the
  arctangent function,'' \emph{IEEE Signal Processing Magazine}, vol.~23,
  no.~3, pp. 187--198, 2006.

\bibitem{rod2018vnni}
A.~Rodriguez, E.~Sen, E.~Meiri, E.~Fomenko, Y.~Jin, H.~Shen, and Z.~Barukh,
  ``Lower numerical precision deep learning inference and training,'' 2018.

\bibitem{1970A&A.....5...84Z}
Y.~B. {Zel'dovich}, ``{Gravitational instability: An approximate theory for
  large density perturbations.}'' \emph{\aap}, vol.~5, pp. 84--89, Mar. 1970.

\bibitem{2017RAA....17...85E}
J.~D. {Emberson}, H.-R. {Yu}, D.~{Inman}, T.-J. {Zhang}, U.-L. {Pen},
  J.~{Harnois-D{\'e}raps}, S.~{Yuan}, H.-Y. {Teng}, H.-M. {Zhu}, X.~{Chen}, and
  Z.-Z. {Xing}, ``{Cosmological neutrino simulations at extreme scale},''
  \emph{Research in Astronomy and Astrophysics}, vol.~17, p. 085, Aug. 2017.

\bibitem{2011JCAP...07..034B}
D.~{Blas}, J.~{Lesgourgues}, and T.~{Tram}, ``{The Cosmic Linear Anisotropy
  Solving System (CLASS). Part II: Approximation schemes},'' \emph{J. Cosmology
  Astropart. Phys.}, vol.~7, p. 034, Jul. 2011.

\bibitem{Smith_2003}
\BIBentryALTinterwordspacing
R.~E. Smith, J.~A. Peacock, A.~Jenkins, S.~D.~M. White, C.~S. Frenk, F.~R.
  Pearce, P.~A. Thomas, G.~Efstathiou, and H.~M.~P. Couchman, ``Stable
  clustering, the halo model and non-linear cosmological power spectra,''
  \emph{Monthly Notices of the Royal Astronomical Society}, vol. 341, no.~4, p.
  1311–1332, Jun 2003. [Online]. Available:
  \url{http://dx.doi.org/10.1046/j.1365-8711.2003.06503.x}
\BIBentrySTDinterwordspacing

\bibitem{ishiyama20124.45}
T.~Ishiyama, K.~Nitadori, and J.~Makino, ``4.45 pflops astrophysical n -body
  simulation on k computer: the gravitational trillion-body problem,'' pp.
  1--10, 2012.

\bibitem{habib2012the}
S.~Habib, V.~Morozov, H.~Finkel, A.~C. Pope, K.~Heitmann, K.~Kumaran,
  T.~Peterka, J.~Insley, D.~Daniel, P.~Fasel \emph{et~al.}, ``The universe at
  extreme scale: Multi-petaflop sky simulation on the bg/q,'' \emph{arXiv:
  Distributed, Parallel, and Cluster Computing}, 2012.

\bibitem{skillman2014}
S.~W. {Skillman}, M.~S. {Warren}, M.~J. {Turk}, R.~H. {Wechsler}, D.~E. {Holz},
  and P.~M. {Sutter}, ``{Dark Sky Simulations: Early Data Release},''
  \emph{arXiv e-prints}, p. arXiv:1407.2600, Jul 2014.

\bibitem{ishiyama2015}
T.~{Ishiyama}, M.~{Enoki}, M.~A.~R. {Kobayashi}, R.~{Makiya}, M.~{Nagashima},
  and T.~{Oogi}, ``{The {\ensuremath{\nu}}$^{2}$GC simulations: Quantifying the
  dark side of the universe in the Planck cosmology},'' \emph{Publications of
  the ASJ}, vol.~67, no.~4, p.~61, Aug 2015.

\bibitem{heitmann2015the}
K.~Heitmann, N.~Frontiere, C.~Sewell, S.~Habib, A.~Pope, H.~Finkel, S.~Rizzi,
  J.~Insley, and S.~Bhattacharya, ``The q continuum simulation: Harnessing the
  power of gpu accelerated supercomputers,'' \emph{Astrophysical Journal
  Supplement Series}, vol. 219, no.~2, p.~34, 2015.

\bibitem{potter2017}
D.~{Potter}, J.~{Stadel}, and R.~{Teyssier}, ``{PKDGRAV3: beyond trillion
  particle cosmological simulations for the next era of galaxy surveys},''
  \emph{Computational Astrophysics and Cosmology}, vol.~4, no.~1, p.~2, May
  2017.

\bibitem{heitmann2019}
K.~{Heitmann}, H.~{Finkel}, A.~{Pope}, V.~{Morozov}, N.~{Frontiere},
  S.~{Habib}, E.~{Rangel}, T.~{Uram}, D.~{Korytov}, H.~{Child}, S.~{Flender},
  J.~{Insley}, and S.~{Rizzi}, ``{The Outer Rim Simulation: A Path to Many-core
  Supercomputers},'' \emph{Astrophysical Journal Supplement Series}, vol. 245,
  no.~1, p.~16, Nov 2019.

\end{thebibliography}

\end{document}